\documentclass{paper}

\renewcommand{\d}{\mathrm{d}}
\begin{document}

\title{Power spectrum from weak-shear data}
\author{Matthias Bartelmann and Peter Schneider\\
  Max-Planck-Institut f\"ur Astrophysik, P.O.~Box 1523, D--85740
  Garching, Germany}

\date{\em Astronomy \& Astrophysics (1999)}

\begin{abstract}
We demonstrate that the aperture mass as a measure for cosmic shear
closely approximates (to better than $\approx5\%$) the scaled and
shifted power spectrum of the projected mass density. This
cosmological weak-lensing information can thus be used to directly
infer the projected matter power spectrum with high accuracy. As an
application, we show that aperture-mass observations can be used to
constrain the cosmic density parameter and the power-spectrum
amplitude. We show that, for a particular example, it should be
possible to constrain $\Omega_0$ to within $\approx\pm27\%$, and
$\sigma_8$ to within $\approx\pm8\%$ using weak-shear data on a
square-shaped field of $8^\circ$ side length.
\end{abstract}

\maketitle

\section{Introduction}

A new measure for cosmic shear, the aperture mass $\langle
M_\mathrm{ap}\rangle$, was recently proposed by Schneider et
al.~(1998; hereafter S98). It was shown there that $\langle
M^2_\mathrm{ap}\rangle$ is related to the power spectrum $P_\kappa$ of
the projected density fluctuations filtered with a narrow
function. Here, we demonstrate in Sect.~2 that an accurate
approximation to $\langle M^2_\mathrm{ap}\rangle$ can be constructed
which directly yields $P_\kappa$. In essence, this makes the projected
matter power spectrum a directly observable quantity, so that no
deconvolution algorithms need to be invoked. We then use our
approximation in Sect.~3 to infer the cosmic density parameter
$\Omega_0$ and the power-spectrum normalisation $\sigma_8$ from
simulated data. We present our conclusions in Sect.~4.

\section{Aperture mass and approximations}

\subsection{Effective convergence}

The two-point statistics of the gravitational lens properties of the
large-scale structure can be described, to high accuracy, in terms of
the power spectrum of an equivalent single lens plane matter
distribution (see, e.g., Kaiser 1998; S98; and references therein),
which is given by
\begin{equation}
  P_\kappa(l) = \frac{9H_0^4\Omega_0^2}{4c^4}\,
  \int_0^{w_\mathrm{H}}\d w\frac{W^2(w)}{a^2(w)}\,
  P_\delta\left(\frac{l}{f_K(w)},w\right)\;,
\label{eq:1}
\end{equation}
where $\vec l$ is the Fourier conjugate to the angle $\vec\theta$, $w$
and $f_K(w)$ are the comoving radial and angular-diameter distance,
respectively, $a=(1+z)^{-1}$ is the scale factor, $P_\delta(k,w)$ is
the density-perturbation power spectrum, and $W(w)$ is the weight
function
\begin{equation}
  W(w) = \int_w^{w_\mathrm{H}}\d w'G(w')\frac{f_K(w'-w)}{f_K(w')}\;,
\label{eq:2}
\end{equation}
which depends on the probability distribution $G(w)$ of source
distances. The upper integration limit $w_\mathrm{H}$ is the comoving
horizon distance, here defined as the comoving distance corresponding
to redshift infinity. Quite intuitively, eq.~(\ref{eq:1}) relates the
lensing power on angular scales $\theta=2\pi l^{-1}$ to the power in
density fluctuations at a comoving scale $2\pi k^{-1}=f_K(w)\theta$.

We parameterise the source-distance distribution $G(w)$ as a function
of redshift, $G_z(z)$, specified by
\begin{equation}
  G_z(z) = \frac{\beta}{z_0^3\Gamma(3/\beta)}\,z^2\,
  \exp\left[-\left(\frac{z}{z_0}\right)^\beta\right]\;.
\label{eq:2a}
\end{equation}
It is normalised to $0\le z<\infty$ and provides a good fit to the
observed redshift distribution (e.g.~Smail et al.~1995). The mean
redshift $\langle z\rangle$ is proportional to $z_0$. For $\beta=1.5$
which we assume throughout, $\langle z\rangle\approx1.505\,z_0$.

\subsection{Aperture mass}

The aperture mass $M_\mathrm{ap}(\theta)$ as a function of smoothing
scale $\theta$ is defined in terms of a weighted average within a
circle of radius $\theta$ of the surface mass density of the
equivalent single lens plane. It can readily be obtained from the
observed image ellipticities of faint background galaxies which
provide an unbiased estimate of the shear $\gamma$,
\begin{equation}
  M_\mathrm{ap}(\theta)=\int\d^2\vartheta\,
  Q(|\vec\vartheta|)\,\gamma_{\mathrm t}(\vec\vartheta)\;,
\label{eq:51}
\end{equation}
by replacing the integral over the shear by a sum over galaxy
ellipticities. Here, $\gamma_\mathrm{t}$ is the tangential component
of the shear relative to the aperture centre, and $Q$ is an
appropriately chosen weight function which is non-zero only for $0\le
\vartheta\le\theta$. The mean-squared aperture mass is related to the
effective-convergence power spectrum through
\begin{equation}
  \langle M^2_\mathrm{ap}\rangle(\theta) =
  2\pi\int_0^\infty\d l\,l\,P_\kappa(l)\,J^2(l\theta)\;,
\label{eq:3}
\end{equation}
where $J(\eta)$ is related to the filter function $Q$ (see S98). For
\begin{equation}
  Q(\vartheta)=\frac{6}{\pi}
  \frac{\vartheta^2(\theta^2-\vartheta^2)}{\theta^6}
  \quad\Rightarrow\quad
  J(\eta) = \frac{12}{\pi\eta^2}\,\mathrm{J}_4(\eta)\;,
\label{eq:4}
\end{equation}
and $\mathrm{J}_4(\eta)$ is the fourth-order Bessel function of the
first kind. $J^2(\eta)$ peaks at $\eta'_0\approx4.11$. Examples for
the {\em rms\/} aperture mass $\langle M^2_\mathrm{ap}\rangle^{1/2}$
are plotted in Fig.~\ref{fig:2} as a function of aperture radius
$\theta$.

It is crucial in eqs.~(\ref{eq:1}) and (\ref{eq:3}) to take the
non-linear evolution of the density power spectrum into account. For
aperture radii of order a few arc minutes,
$\theta\approx3\times10^{-4}\,\mathrm{rad}$, the peak $\eta'_0$ of
$J^2(\eta)$ translates to $l\approx1.4\times10^4$, which corresponds
to $2\pi k^{-1}\approx1\,h^{-1}\mathrm{Mpc}$ for sources around
redshift unity, i.e.~the physical scales of density perturbations to
which the aperture mass is most sensitive to are well in the
non-linear regime of evolution. We assume a CDM power spectrum and
describe its non-linear evolution as given by Peacock \& Dodds
(1996). We normalise the spectrum to the local abundance of rich
clusters by choosing $\sigma_8$ as derived by Viana \& Liddle (1996)
and Eke, Cole \& Frenk (1996).

\subsection{Signal-to-noise ratio}

Although an {\em rms\/} aperture-mass amplitude around one per cent
appears low, the signal-to-noise ratio of the aperture mass can be
quite high. There are three sources of noise in a measurement of
$\langle M^2_\mathrm{ap}\rangle$. Since it will be inferred from
distortions of galaxy ellipticities, the intrinsic non-vanishing
ellipticity of the galaxies provides one source of noise, and the
random positions of the galaxies provides another. The third source of
noise is due to cosmic variance. Assuming a large number of galaxies
$N$ per aperture, and neglecting the kurtosis of the aperture mass,
the dispersion of $M_\mathrm{ap}$ in a {\em single\/} aperture is
(see S98)
\begin{equation}
  \sigma^2(M^2_\mathrm{ap}) \approx \left(
    \frac{6\sigma^2_\epsilon}{5\sqrt{2}N} +
    \sqrt{2}\langle M^2_\mathrm{ap}\rangle
  \right)^2\;,
\label{eq:5}
\end{equation}
where $\sigma_\epsilon\approx0.2$ is the intrinsic dispersion of the
galaxy ellipticities. On angular scales of a few arc minutes and
smaller, the galaxies dominate the noise, while the cosmic variance
dominates on larger scales. Of course, $M_\mathrm{ap}$ will be
measured in a large number of apertures $N_\mathrm{ap}$ rather than a
single one. If the apertures are independent, the dispersion
(\ref{eq:5}) is reduced by a factor of $N_\mathrm{ap}^{1/2}$, and the
ensemble variance becomes
\begin{equation}
  \bar{\sigma} =
  \frac{\sigma(M^2_\mathrm{ap})}{N^{1/2}_\mathrm{ap}}\;.
\label{eq:6}
\end{equation}
Typical signal-to-noise ratios reach values of $\gtrsim5$ for aperture
radii of a few arc minutes and data fields of a square degree in size.

\subsection{Approximate {\em rms\/} aperture mass}

The strong peak of $J^2(\eta)$ -- see Fig.~2 in S98 -- motivates the
approximation of $J^2$ by a delta function,
\begin{equation}
  J^2(\eta)\approx A\,\delta_\mathrm{D}(\eta-\eta_0)\;,
\label{eq:8}
\end{equation}
with $A=512/(1155\pi^3)\approx1.43\times 10^{-2}$,
$\eta_0=693\pi/512\approx4.25$, as determined from the norm and mean
of $J^2$.\footnote{In fact, the peak of $J^2(\eta)$ can well be
approximated by a Gaussian. A least-square fit of a Gaussian to
$J^2(\eta)$ yields
\begin{eqnarray}
  J^2(\eta) &\approx&
  A'\,\exp\left[-(\eta-\eta'_0)^2/(2\sigma^2)\right] \nonumber
\end{eqnarray}
with mean $\eta'_0\approx4.11$, amplitude
$A'\approx4.52\times10^{-3}$, and variance $\sigma\approx1.24$. The
small width of the Gaussian encourages the approximation
(\ref{eq:8}).} Then the expression for the aperture mass (\ref{eq:3})
becomes particularly simple,
\begin{equation}
  \langle M^2_\mathrm{ap}\rangle(\theta) \approx
  \langle\tilde{M}^2_\mathrm{ap}\rangle(\theta) \equiv
  6 (5\pi)^{-1}\,\theta^{-2}\,P_\kappa(\eta_0/\theta)\;.
\label{eq:9}
\end{equation}
If this provides a good approximation, the observable {\em rms\/}
aperture mass at angular scale $\theta$ would directly yield the
effective-convergence power spectrum $P_\kappa(l)$ at wave number
$\eta_0/\theta$, and thus a most direct and straightforward measure
for dark-matter fluctuations. We therefore have to investigate whether
replacing the filter function $J^2(\eta)$ by a delta function can be
justified.

\begin{figure}[ht]
  \includegraphics[width=\hsize]{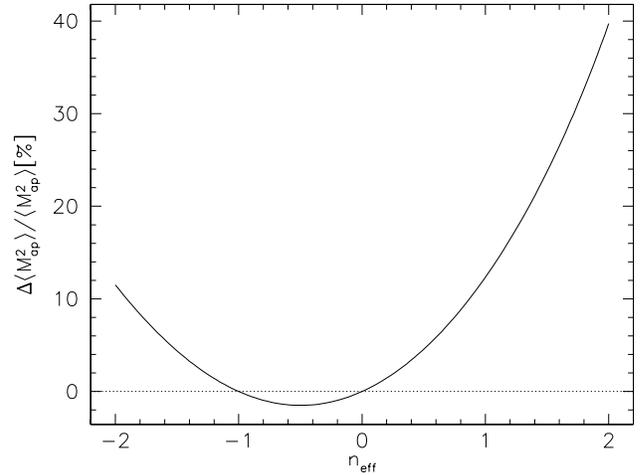}
\caption{Relative deviation between the true aperture mass $\langle
  M^2_\mathrm{ap}\rangle$ and the approximate aperture mass
  $\langle\tilde{M}^2_\mathrm{ap}\rangle$ as defined in
  eq.~(\ref{eq:9}), for an assumed effective-convergence power
  spectrum that is a power law in $l$. For the most relevant range of
  effective power-law exponents, $-1.5\lesssim n_\mathrm{eff}\lesssim
  0.5$, the relative deviation is less than five per cent.}
\label{fig:1}
\end{figure}

\begin{figure}[ht]
  \includegraphics[width=\hsize]{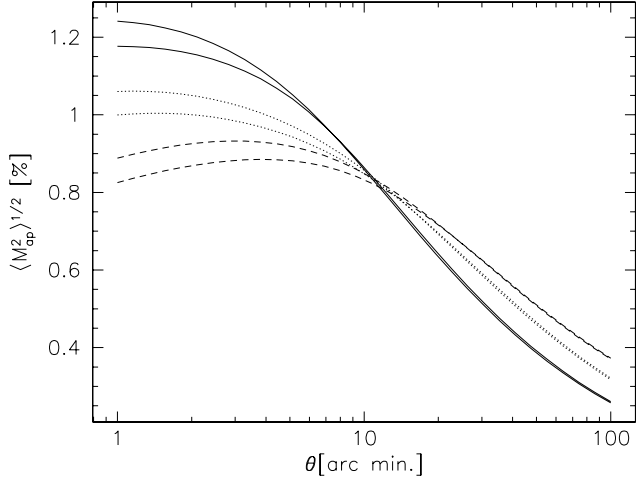}
\caption{Exact and approximate {\em rms\/} aperture masses $\langle
  M^2_\mathrm{ap}\rangle$ and $\langle\tilde{M}^2_\mathrm{ap}\rangle$
  are shown for three cosmological models. All are cluster-normalised
  CDM models in which the non-linear growth of the density power
  spectrum $P_\delta$ was taken into account. Solid curves:
  low-density, open model ($\Omega_0=0.3$, $\Omega_\Lambda=0$,
  $h=0.7$); dotted curves: low-density, spatially flat model
  ($\Omega_0=0.3$, $\Omega_\Lambda=0.7$, $h=0.7$); dashed curves:
  Einstein-de Sitter model ($\Omega_0=1$, $\Omega_\Lambda=0$,
  $h=0.5$). The redshift distribution (\ref{eq:2a}) of the sources was
  used. Typical {\em rms\/} aperture masses reach $\sim 1$ per cent at
  angular scales of a few arc minutes. The normalisation to the local
  abundance of rich clusters causes the curves to be very similar for
  the different choices of the cosmological parameters.  While the
  approximate aperture mass slightly underestimates the true aperture
  mass at small aperture radii, $\theta\gtrsim1'$, by at most
  $\approx5\%$, the approximations fall exactly on top of the true
  curves for angular scales beyond $\approx10'$.}
\label{fig:2}
\end{figure}

Let us first assume that $P_\kappa(l)$ can be approximated locally as
a power law in $l$ over a range in which $J^2$ differs significantly
from zero,
\begin{equation}
  P_\kappa(l) = B\,l^{n_\mathrm{eff}}\;.
\label{eq:10}
\end{equation}
Then, the true aperture mass (\ref{eq:3}) becomes
\begin{equation}
  \langle M^2_\mathrm{ap}\rangle(\theta) =
  \frac{144B}{\pi^{3/2}}\,
  \frac{
    \Gamma\left(\frac{3-n_\mathrm{eff}}{2}\right)
    \Gamma\left(3+\frac{n_\mathrm{eff}}{2}\right)
  }{
    \Gamma\left(2-\frac{n_\mathrm{eff}}{2}\right)
    \Gamma\left(6-\frac{n_\mathrm{eff}}{2}\right)
  }\,\theta^{-(2+n_\mathrm{eff})}\;,
\label{eq:11}
\end{equation}
where the Gamma functions arise from integrating the power-law
spectrum times the squared fourth-order Bessel function. The
approximate aperture mass $\langle\tilde{M}^2_\mathrm{ap}\rangle$
defined in eq.~(\ref{eq:9}) becomes
\begin{equation}
  \langle\tilde{M}^2_\mathrm{ap}\rangle(\theta) =
  6(5\pi)^{-1} B\,\eta_0^{n_\mathrm{eff}}\,
  \theta^{-(2+n_\mathrm{eff})}\;.
\label{eq:12}
\end{equation}
The relative deviation of the approximation from the true aperture
mass is a function of the effective power-law exponent
$n_\mathrm{eff}$ only. It is plotted in Fig.~\ref{fig:1}. Clearly, the
deviation of $\langle M^2_\mathrm{ap}\rangle$ from
$\langle\tilde{M}^2_\mathrm{ap}\rangle$ is very small, less than five
per cent, for effective power-spectrum slopes in the range
$-1.5\lesssim n_\mathrm{eff}\lesssim0.5$, and this is exactly the
range of slopes in the $l$ interval contributing most of the power to
the aperture mass for aperture radii of $\gtrsim1'$. It is therefore
fair to say that, especially in the presence of measurement errors,
the aperture mass and its approximation (\ref{eq:9}) can be considered
equivalent. Figure~\ref{fig:2} shows $\langle M^2_\mathrm{ap}\rangle$
and $\langle\tilde{M}^2_\mathrm{ap}\rangle$ for three
cluster-normalised CDM models as examples for realistic non-power-law
spectra, and emphasises the conclusions from Fig.~\ref{fig:1}. The
approximation (\ref{eq:9}) to the aperture mass is excellent. This
immediately implies that measurements of the aperture mass directly
measure the effective-convergence power spectrum, and therefore the
latter effectively becomes an observable quantity,
\begin{equation}
  P_\kappa(l)\approx (5\pi/6)\,(\eta_0/l)^2\,\langle
  M^2_\mathrm{ap}\rangle(\eta_0/l) \;.
\label{eq:52}
\end{equation} 
This provides the most straightforward access to a quantity of
paramount importance for cosmology, i.e.~the dark-matter power
spectrum.

\section{Applications}

\begin{figure}[ht]
  \includegraphics[width=\hsize]{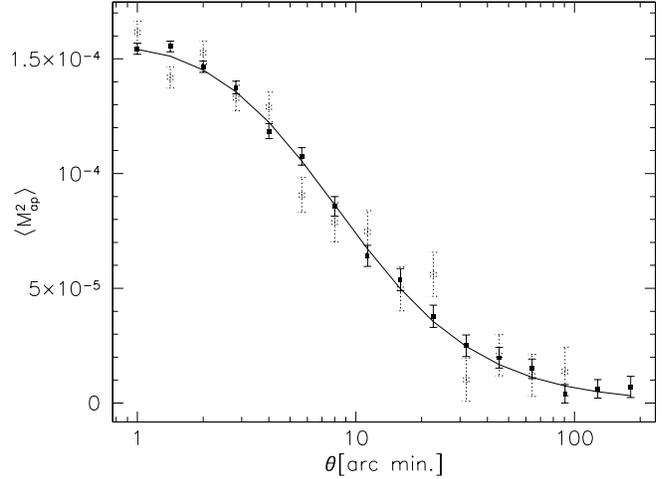}
\caption{Simulated aperture-mass measurements employing the sampling
  strategy detailed in the text. The error bars show 1-$\sigma$ errors
  derived from eq.~(\ref{eq:6}). Field sizes of $4^\circ$ and
  $8^\circ$ were assumed for the open squares with dotted error bars
  and the filled squares with solid error bars, respectively. The true
  aperture mass is shown as the solid curve. The low-density, open CDM
  model was assumed for this plot, and $z_0=1$ was chosen for the
  redshift distribution of the sources. According to eq.~(\ref{eq:9}),
  the effective-convergence power spectrum can directly be read off
  from this plot.}
\label{fig:3}
\end{figure}

It is now interesting to investigate a specific application of our
results.  Let us assume we are given a large, quadratic data field of
angular size $L$, into which we place apertures of radius $\theta$. We
can place apertures of a fixed size densely on the field since the
correlation of the aperture mass in neighbouring apertures is very
small due to the narrowness of the filter function $J^2(\eta)$ (see
Fig.~8 in S98). However, we want to sample $\langle
M^2_\mathrm{ap}\rangle$ at a variety of aperture sizes rather than a
single one. Starting from a smallest aperture radius $\theta_0$, we
choose the next largest aperture radius as $\sqrt{2}\theta_0$ and so
forth, so that consecutive aperture radii encompass on average twice
as many galaxies. Clearly, the data for different aperture sizes are
not completely uncorrelated, but decorrelate quickly with differing
radius and centre separation -- see again Fig.~8 of S98. Therefore, we
assume that the number of independent apertures of radius $\theta$ is
given by $N_\mathrm{ap}=[L/(2\theta)]^2$, and the expected number of
galaxies per aperture is $N=n\pi\theta^2$. We assume a galaxy density
of $n=30$ per square arc minute. The expectation value of $\langle
M^2_\mathrm{ap}\rangle(\theta)$ is given by eq.~(\ref{eq:3}). Its
1-$\sigma$ error at fixed aperture radius is expected to be given by
eq.~(\ref{eq:6}). We show examples of simulated measurements $M_i$
together with their expected 1-$\sigma$ error bars in Fig.~\ref{fig:3}
for field sizes of $4^\circ$ and $8^\circ$.

\begin{figure}[ht]
  \includegraphics[width=\hsize]{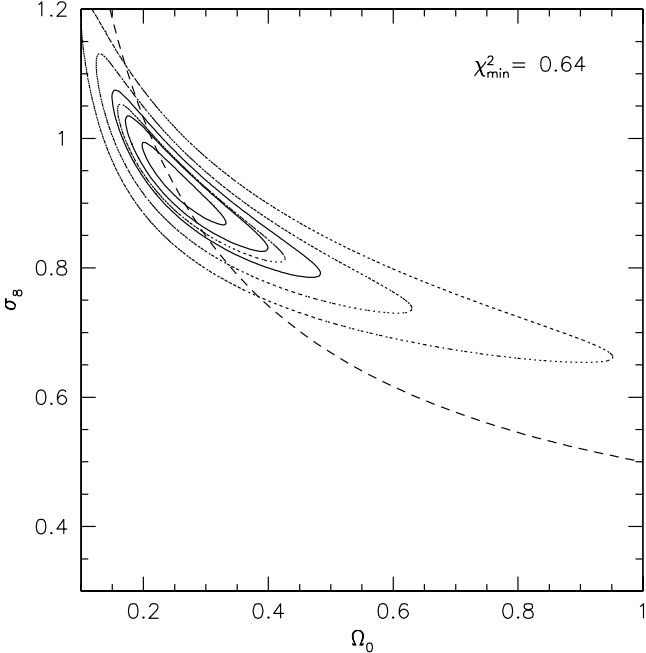}
\caption{Contours of $\chi^2$ as defined in eq.~(\ref{eq:13}) are
  shown in the $\Omega_0-\sigma_8$ plane. Input data and their errors
  were simulated as detailed in the text, assuming quadratic data
  fields of side length $4^\circ$ and $8^\circ$ (dotted and solid
  contours, respectively). Contours are shown for 1, 2, and 3-$\sigma$
  significance levels. The input cosmological model was the
  cluster-normalised, low-density, open CDM model with $\Omega_0=0.3$,
  $\Omega_\Lambda=0$ and $h=0.7$. Due to the cluster normalisation,
  $\sigma_8=0.85$. $\Omega_\Lambda=0$ was assumed for the fit.  The
  contours show that data on a $8^\circ\times8^\circ$ field are
  sufficient to constrain $\Omega_0$ at the 1-$\sigma$ level to within
  $0.18\lesssim\Omega_0\lesssim0.34$, and the amplitude to
  $0.85\lesssim\sigma_8\lesssim0.99$. The long-dashed curve shows the
  present-day cluster-abundance constraint on $\Omega_0$ and
  $\sigma_8$.}
\label{fig:4}
\end{figure}

Let $M_i$ be the measurement of $\langle M^2_\mathrm{ap}\rangle$ for
aperture radius $\theta_i$, then (\ref{eq:6}) yields the expected
error of $M_i$. We can then try and fit the ensemble of measurements
with the approximate aperture mass
$\langle\tilde{M}^2_\mathrm{ap}\rangle$. For that purpose, we define
the $\chi^2$ function
\begin{equation}
  \chi^2 = \frac{1}{N_\theta}\sum_{i=1}^{N_\theta}\,
  \frac{[M_i-\langle\tilde{M}^2_\mathrm{ap}\rangle(\theta_i)]^2}
       {\bar\sigma_i^2}\;.
\label{eq:13}
\end{equation}
As an example, we assume that data points like those shown in
Fig.~\ref{fig:3} have been measured in a cluster-normalised,
low-density open CDM model. Assuming the CDM shape of the
density-perturbation power spectrum, we can then vary parameters such
as to minimise (\ref{eq:13}). The sensitivity of the aperture mass to
the Hubble constant is very small. The remaining parameters are then
the parameters $\beta$ and $z_0$ of the source redshift distribution,
the amplitude of the power spectrum, parameterised by $\sigma_8$, the
density parameter $\Omega_0$, and the cosmological constant
$\Omega_\Lambda$. It can safely be assumed that the redshift
distribution $G(w)$ of background galaxies is known with high accuracy
from other observations. We can therefore restrict ourselves to
essentially two free parameters, $\sigma_8$ and $\Omega_0$, and choose
either $\Omega_\Lambda=0$ or $\Omega_\Lambda=1-\Omega_0$. $\chi^2$
contours for the former case are shown in Fig.~\ref{fig:4}.

The solid contours in the left panel of Fig.~\ref{fig:4} show that at
the 1-$\sigma$ significance level, $\Omega_0$ and $\sigma_8$ can be
constrained to relative accuracies of $\approx\pm27\%$ and
$\approx\pm8\%$, respectively. Table~\ref{tab:1} gives 1-$\sigma$
limits for fields of $4^\circ$ and $8^\circ$ side length. Similar
experiments using $\Omega_\Lambda=1-\Omega_0$ instead of
$\Omega_\Lambda=0$ yield a minimum $\chi^2\gtrsim10$, so that
spatially flat cosmologies can be rejected at very high significance
using the data simulated with $\Omega_\Lambda=0$. 

Of course, the quality of the results depends on how accurately we
guess the shape of the true power spectrum. For the contours in
Fig.~\ref{fig:4}, we vary two parameters and keep all others fixed. In
Fig.~\ref{fig:4} for example, we keep $h=0.7$ constant, so that the
shape parameter $\Gamma\equiv\Omega_0\,h$ of the power spectrum is
$\Gamma=0.7\Omega_0$. If we would set the shape parameter instead to
$\Gamma'=f\Gamma$ with $f\ne1$, the best-fitting parameter combination
($\Omega_0$, $\sigma_8$) would shift towards $f\Omega_0$ along the
``valley'' indicated by the $\chi^2$ contours without substantially
degrading the quality of the fit. This degeneracy reflects the
degeneracy in the cluster abundance used to normalise the power
spectrum for simulating the input data, because $\langle
M_\mathrm{ap}^2\rangle^{1/2}$ essentially measures the halo abundance
at intermediate redshifts, $z\sim0.3$. To illustrate this point, we
indicate the present-day cluster-abundance constraint as the
long-dashed curve in Fig.~\ref{fig:4}. It is seen to follow well the
``valley'' in the $\chi^2$ contours for low $\Omega_0$, but it departs
increasingly for increasing $\Omega_0$ because structure grows more
rapidly between $z\sim0.3$ and today for higher $\Omega_0$. The
parameter degeneracy with the cosmological constant can be broken
using the skewness of $M_\mathrm{ap}$, for which the degeneracy
between $\Omega_0$ and $\Omega_\Lambda$ is different (Van Waerbeke,
Bernardeau \& Mellier 1999).

Although our assumptions concerning the number of independent
apertures that can be placed inside a given data field and the neglect
of the kurtosis in the approximation for $\sigma(M_{\mathrm ap}^2)$
may be slightly optimistic, we believe that these effects do not
change our conclusions appreciably. It should be noted that for
applying our technique, the data field need not be a connected region,
but arbitrarily placed smaller areas on the sky with about equal total
solid angle are equally useful. The rapid evolution towards
square-degree CCD cameras at the best observing sites leaves us highly
optimistic concerning the future application of our method.

\begin{table}[ht]
\caption{1-$\sigma$ limits to $\Omega_0$ and $\sigma_8$ obtained with
  fields of $4^\circ$ and $8^\circ$ side length.}
\label{tab:1}
\begin{center}
\begin{tabular}{|r|rr|rr|}
\hline
  & \multicolumn{2}{|c}{$\Omega_0$} &
    \multicolumn{2}{|c|}{$\sigma_8$} \\
  $L$ & input & fit & input & fit \\
\hline
  $4^\circ$ & $0.3$ & $0.25\pm0.12$ & $0.85$ & $0.93\pm0.12$ \\
  $8^\circ$ & $0.3$ & $0.26\pm0.08$ & $0.85$ & $0.93\pm0.08$ \\
\hline
\end{tabular}
\end{center}
\end{table}

\section{Conclusions}

We have shown here that a recently proposed measure for cosmic shear,
the dispersion of the aperture mass $\langle
M_\mathrm{ap}^2(\theta)\rangle$, can be approximated by the local
value of the power spectrum $P_\kappa$ of the projected matter
fluctuations at wave number $\eta_0/\theta$ to a relative accuracy
better than $5\%$. This is possible because the aperture mass is a
convolution of $P_\kappa$ with a narrow filter function which for this
purpose can safely be approximated by a Dirac delta function. 

In contrast to other methods for constraining $P_\kappa(l)$ through
cosmic-shear data (Kaiser 1998; Seljak 1998; Van Waerbeke et
al.~1999), the aperture-mass method allows for a simple combined
analysis of independent data fields scattered across the
sky. Furthermore, since $\langle M_\mathrm{ap}^2\rangle^{1/2}$ is a
scalar quantity directly obtained from observed galaxy-image
ellipticities, its full probability-distribution function can be
derived and used for parameter extraction.

The approximate aperture mass can then straightforwardly be applied to
constrain cosmological parameters from observed aperture masses. We
simulated observations of $\langle M^2_\mathrm{ap}\rangle$ and their
expected errors and showed that $\Omega_0$ and $\sigma_8$ can be
recovered with relative accuracies of $\approx\pm27\%$ and
$\approx\pm8\%$, respectively, using weak-shear data on a square field
with $8^\circ$ side length. For that, we have assumed that the
parameters of the source redshift distribution are sufficiently well
known. We believe that this is not a serious limitation because of the
high reliability and accuracy of photometric redshift determinations
(e.g., Ben{\'\i}tez 1998, and references therein). In fact, the
source-redshift dependence of $\langle M^2_\mathrm{ap}\rangle$ can be
used as a consistency check: With increasing source redshift, $\langle
M^2_\mathrm{ap}\rangle$ {\em increases\/}, while the skewness of
$M_\mathrm{ap}$ {\em decreases\/}.

\section*{Acknowledgements}

We wish to thank the referee, Ludovic Van Waerbeke, for his detailed
and constructive comments. This work was supported in part by the
Sonderforschungsbereich 375 on Astro-Particle Physics of the Deutsche
Forschungsgemeinschaft.


\begin{thebibliography}{99}

\bibitem{ref:1} Ben{\'\i}tez, N., 1998, preprint astro-ph/9811189
\bibitem{ref:2} Eke, V.R., Cole, S., Frenk, C.S., 1996, MNRAS, 282,
  263
\bibitem{ref:3} Kaiser, N., 1998, ApJ, 498, 26
\bibitem{ref:4} Peacock, J.A., Dodds, S.J., 1996, MNRAS, 280, L19
\bibitem{ref:5} Schneider, P., van Waerbeke, L., Jain, B., Kruse, G.,
  1998, MNRAS, 296, 873 (S98)
\bibitem{ref:6} Seljak, U., 1998, ApJ, 506, 64
\bibitem{ref:7} Smail, I., Hogg, D.W., Yan, L., Cohen, J.G., 1995,
  ApJ, 449, L105
\bibitem{ref:9} Van Waerbeke, L., Bernardeau, F., Mellier, Y., 1999,
  A\&A, 342, 15
\bibitem{ref:8} Viana, P.T.P., Liddle, A.R., 1996, MNRAS, 281, 323

\end{thebibliography}
\end{document}